\documentclass[nofootinbib,prl,twocolumn]{revtex4}

\usepackage{amsmath}
\usepackage{amsthm}
\usepackage{blindtext}
\usepackage{algorithm}
\usepackage{algorithmic}
\usepackage{amssymb}
\usepackage{graphicx}
\usepackage{color}
\usepackage{array}
\usepackage[makeroom]{cancel}
\usepackage{bbm}
\usepackage[authormarkup=none]{changes}
\usepackage{changes}
\usepackage{algorithm}
\usepackage[hidelinks]{hyperref}

\DeclareMathOperator{\id}{\mathbbm{1}}

\DeclareMathOperator{\Haf}{\text{Haf}}

\newcommand{\beq}{\begin{equation}}
\newcommand{\eeq}{\end{equation}}

\begin{document}
\title{Using Gaussian Boson Sampling to Find Dense Subgraphs}
\author{Juan Miguel Arrazola}
\email{juanmiguel@xanadu.ai}
\author{Thomas R. Bromley}
\email{tom@xanadu.ai}
\affiliation{Xanadu, 372 Richmond Street W, Toronto, Ontario M5V 1X6, Canada}

\begin{abstract}
Boson sampling devices are a prime candidate for exhibiting quantum supremacy, yet their application for solving problems of practical interest is less well understood. Here we show that Gaussian boson sampling (GBS) can be used for dense subgraph identification. Focusing on the NP-hard densest $k$-subgraph problem, we find that stochastic algorithms are enhanced through GBS, which selects dense subgraphs with high probability. These findings rely on a link between graph density and the number of perfect matchings -- enumerated by the Hafnian -- which is the relevant quantity determining sampling probabilities in GBS. We test our findings by constructing GBS-enhanced versions of the random search and simulated annealing algorithms and apply them through numerical simulations of GBS to identify the densest subgraph of a $30$ vertex graph. 
\end{abstract}

\maketitle

Quantum algorithms are often designed with the assumption that they can access the full power of universal quantum computation. However, presently developing quantum devices have limited resource capabilities and are not fault-tolerant. Their emergence has motivated a reexamination of methods for designing quantum algorithms, with the focus now on harnessing the computational power of small-scale, noisy quantum computers. Candidate algorithms for near-term devices include quantum simulators for many-body physics \cite{zhang2017observation, bernien2017probing}, variational algorithms \cite{peruzzo2014variational,mcclean2016theory,moll2017quantum, kandala2017hardware}, quantum approximate optimization algorithms \cite{farhi2014quantum, farhi2016quantum}, and machine learning on hybrid devices \cite{li2015experimental,riste2017demonstration,benedetti2017quantum,otterbach2017unsupervised,schuld2018quantum}.

Boson sampling is a limited model of quantum computation given by passing photons through a linear interferometer and observing their output configurations~\cite{aaronson2011computational}. Significant efforts have been performed to implement boson sampling \cite{spring1231692, broome794, tillmann2013experimental, crespi2013integrated}, leading to the proposal of related models such as scattershot boson sampling \cite{lund2014boson,bentivegna2015experimental,latmiral2016towards} and Gaussian boson sampling \cite{hamilton2017gaussian, kruse2018detailed} that are more suitable for experimental realizations. Moreover, boson sampling devices are in principle capable of performing tasks that cannot be efficiently simulated on classical computers, a feature that has made them a leading candidate for challenging the extended Church-Turing thesis. In fact, the primary objective of implementing boson sampling has so far been to demonstrate quantum supremacy, leaving the real-world application of such devices underdeveloped. A notable exception is the use of Gaussian boson sampling for efficiently calculating the vibronic spectra of molecules, \cite{huh2015boson, clements2017experimental, sparrow2018simulating}, which provided the first clue of the usefulness of this platform.

In this work, we show that Gaussian boson sampling (GBS) can be used to enhance classical stochastic algorithms for the densest $k$-subgraph (D$k$S) problem. The D$k$S problem is NP-Hard \cite{feige2001dense} and defined through the following optimization task: given a graph $G$ with $n$ vertices, find the subgraph of $k<n$ vertices with the largest density. Among subgraphs with a fixed number of vertices, the density and the number of edges are equivalent quantities, and we hence refer to both interchangeably throughout this manuscript. Beyond its fundamental interest in mathematics and theoretical computer science, the D$k$S problem has a natural connection to clustering problems with the goal of finding highly correlated subsets of data. Clustering has applications in a wide range of fields such as data mining \cite{kumar1999trawling,angel2012dense,beutel2013copycatch,chen2012dense}, bioinformatics \cite{fratkin2006motifcut,saha2010dense}, and finance \cite{arora2011computational}.

Our approach uses a technique from Ref.~\cite{bradler2017gaussian} to encode a graph into the GBS paradigm. Here, the probability of observing a given photon configuration is proportional to the number of perfect matchings of the corresponding subgraph. We highlight a correspondence between the number of perfect matchings in a subgraph and its density, meaning that a suitably programmed GBS device will prefer to output dense subgraph configurations. Following results in a companion paper~\cite{arrazola2018quantum_published}, we see that this is a form of proportional sampling that can be used to enhance the stochastic element of classical optimization heuristics for the D$k$S problem. Since no polynomial-time approximation schemes are believed to exist for the D$k$S problem~\cite{manurangsi2017almost}, certain worst-case instances requiring superpolynomial runtime may be best tackled using stochastic algorithms. Our findings are illustrated for a fixed graph, where we introduce GBS-enhanced hybridizations of random search and simulated annealing algorithms. This approach highlights a general principle of using output samples from a GBS device to enhance approximate solutions to optimization problems.

\textit{Applying GBS to the D$k$S problem.---} The important concepts of GBS are first briefly reviewed. In GBS, photon-number detection is performed on a multi-mode Gaussian state ~\cite{hamilton2017gaussian,kruse2018detailed,weedbrook2012gaussian}.
For an $n$-mode system, we denote the possible outputs of GBS by vectors $S=(s_1,s_2,\ldots,s_{n})$, where $s_i$ is the number of photons detected in output mode $i$. It was shown in Ref.~\cite{hamilton2017gaussian} that the probability of observing an output pattern $S$ is
\beq\label{Eq: GBS_dbn}
P(S)=\left|\sigma_{Q}\right|^{-\frac{1}{2}}\frac{\Haf(\mathcal{A}_S)}{s_1!s_2!\cdots s_n!},
\eeq
where $\sigma_Q=\sigma + \id_{2n}/2$, $\sigma$ is the $(2n\times2n)$-dimensional covariance matrix of the $n$-mode Gaussian state, and $\mathcal{A}_S$ is a submatrix of $\mathcal{A} = \left(\begin{array}{cc}
0 & \id_{n} \\
\id_{n} & 0
\end{array} \right) \left[ \id_{2n}  - \sigma_{Q}^{-1}\right]$ fixed by $S$.  The function $\Haf(\mathcal{A}_{S})$ is the Hafnian of $\mathcal{A}_{S}$~\cite{barvinok2016combinatorics}.

Following Ref. \cite{bradler2017gaussian}, given the adjacency matrix $A$ of an $n$ vertex graph $G$, we set $\mathcal{A} :=c(A \oplus A)$, where $c< \lambda^{-1}$ and $ \lambda$ is the largest eigenvalue of $A$. The resulting covariance matrix is such that its corresponding Gaussian state is pure and can hence be prepared by injecting single-mode squeezed states into a linear optics interferometer~\cite{weedbrook2012gaussian}. We focus on post-selecting output samples from GBS such that $s_{i} \in \{0, 1\}$ and $\sum_{i}s_{i} = k$ for a fixed even $k$, i.e., the set of samples with even-$k$ photons and where no output mode has more than one photon detected -- referred to here as the $k$ collision-free subspace. The probability of getting such an event from GBS is $p_{k\mbox{cf}} := p(k \land \mbox{cf}) = p(\mbox{cf}|k)p(k)$, where $p(\mbox{cf}|k)$ is the collision-free probability given $k$ photons and $p(k)$ is the probability of $k$ photons. Here, $p(\mbox{cf}|k)$ is fixed by the size of $k$ in comparison to $n$, and is expected to be close to unity for $k \ll n$. On the other hand, $p(k)$ is controlled by the amount of input squeezing and can be maximized by the user through the choice of $c$. 

By post-selecting on the $k$ collision-free subspace, the probability of a valid output pattern $S$ is
\begin{equation}\label{Eq: GBS Graphs}
P_{k\mbox{cf}}(S) = \left|\sigma_{Q}\right|^{-\frac{1}{2}} \frac{c^2|\Haf\left(A_{S}\right)|^2}{p_{k\mbox{cf}}},
\end{equation}
where $A_{S}$ is the adjacency matrix corresponding to the subgraph of $A$ selected by $S$. Crucially, the Hafnian of an adjacency matrix is equal to the number of perfect matchings in the corresponding graph, i.e., the number of independent sets of edges in which every vertex of the graph is connected to exactly one edge \cite{barvinok2016combinatorics}. Equation~(\ref{Eq: GBS Graphs}) hence highlights a remarkable feature: the greater the number of perfect matchings in a subgraph, the more likely its corresponding sample is to be outputted through GBS.

\begin{center}
\begin{figure}[t!]
\includegraphics[width= 0.9\columnwidth]{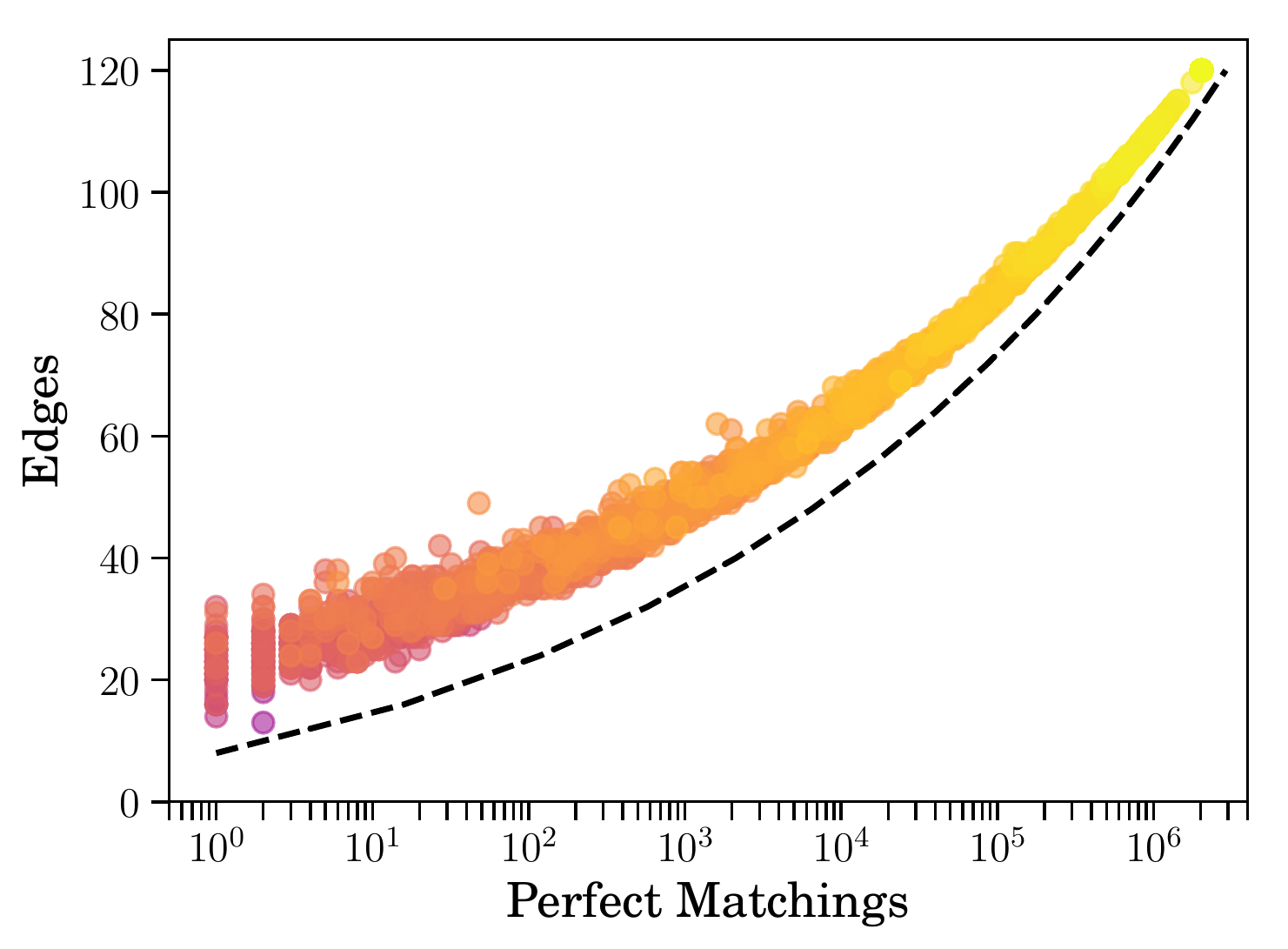}
\caption{Relationship between the number of perfect matchings and the number of edges for random graphs of $k=16$ vertices. The random graphs are generated by adding each possible edge with probability $p$. We generate samples for each value of $p=0.1,0.2,\ldots,1$ for a total of approximately $6000$ random graphs, and plot their number of edges against the Hafnian of the adjacency matrix.  The dashed line is the lower bound from Eq.~\eqref{Eq: PM_bound}.}\label{Fig: Den_Hafnian}
\end{figure}
\end{center}

Our next step is to highlight a correspondence between the number of perfect matchings in a graph and its density. Intuitively, a graph with many perfect matchings is expected to contain many edges. This intuition was made quantitative in Ref. \cite{aaghabali2015upper}, where it was shown that the number of perfect matchings in a graph $G$ with $2m$ vertices is upper bounded by a monotonically increasing function of the number of edges $l$, i.e.,
\beq\label{Eq: PM_bound}
\text{PM}(G)\leq \left(\left\lfloor \frac ln\right\rfloor !\right)^{\frac{m-\alpha}{\lfloor \frac lm\rfloor}}\left(\left\lceil \frac lm\right\rceil !\right)^{\frac{\alpha}{\lceil \frac lm\rceil}},
\eeq
where $\alpha:=l-m\left\lfloor \frac lm\right\rfloor$. Thus, given the number of perfect matchings in a graph with $k:=2m$ vertices, Eq. \eqref{Eq: PM_bound} provides a lower bound to the number of edges in the graph. Fig.~\ref{Fig: Den_Hafnian} illustrates the close relationship between the number of perfect matchings and edges of random graphs, highlighting the usefulness of the above bound. This relationship provides a crucial insight: when sampling from the GBS distribution of Eq.~\eqref{Eq: GBS Graphs}, the subgraphs that are most likely to appear have high density. 

Hence, by programming a GBS device appropriately, it is possible to sample from a distribution that naturally favors dense subgraphs. This is an example of proportional sampling, as described in Ref.~\cite{arrazola2018quantum_published}. In fact, as can be seen in Fig.~\ref{Fig: Den_Hafnian}, the Hafnians of dense graphs can be many orders of magnitude larger than the Hafnians of sparser graphs. For example, the Hafnian of a complete graph of $k$ vertices is equal to $(k-1)!!$. Through proportional sampling, this means that the probability of finding dense graphs is augmented by a correspondingly large factor.  Conversely, graphs with few edges will have either zero or negligible Hafnians, and will therefore almost never be sampled. The combined effect of these features is a GBS distribution that ignores sparse graphs and gives us a much improved chance of discovering the dense ones. 

Proportional sampling leads to a simple algorithm for approximately solving the D$k$S problem for even $k$: generate many samples from GBS with $\mathcal{A} = c(A \oplus A)$ and pick the subgraph with the largest density. For odd $k$, one can output $k+1$ vertex subgraphs and remove the vertex with the lowest degree. This amounts to an enhanced random search algorithm. However, it is often of interest to use more advanced stochastic algorithms that also harness the local structure of an optimization landscape to improve beyond random search. We discuss in the following how simulated annealing can be enhanced for solving the D$k$S problem by using randomness from GBS.

Before doing so, we motivate the use of a physical GBS device for proportional sampling according to Eq.~(\ref{Eq: GBS Graphs}). Indeed, since the Hafnian of an adjacency matrix can be classically approximated in polynomial time~\cite{rudelson2016hafnians}, there exist polynomial-time classical approaches for approximate GBS, such as using rejection sampling or metropolized independent sampling~\cite{neville2017classical,liu1996metropolized}. A physical GBS device, on the other hand, requires constant time to output a sample, leading to a polynomial advantage over these classical methods. Moreover, GBS devices can in principle have very fast sample rates, limited primarily by detector dead times. We also emphasize the inherent robustness of our approach to noise and imperfections in the device, which may typically degrade the intrinsic bias of GBS but not eliminate it completely.

\textit{Enhancing stochastic algorithms through GBS.---} There is a varied collection of classical algorithms for finding dense subgraphs, see for example Ref. \cite{lee2010survey} for a survey. Among these are randomized and deterministic algorithms, each suitable for specific scenarios. Determinstic greedy algorithms can efficiently find subgraphs of large density, but they can be fooled by graphs with special structure. For instance, a widely used algorithm of Charikar~\cite{charikar2000greedy} relies on iteratively eliminating vertices with the lowest degree, but it is incapable of detecting isolated dense subgraphs. On the other hand, the randomness in stochastic algorithms allows them to avoid being fooled by special graph structure, making them a natural choice when little is known about the graph under consideration.  In terms of computational complexity, no polynomial-time approximation scheme exists for solving the D$k$S problem to constant multiplicative error~\cite{manurangsi2017almost} unless the exponential time hypothesis is false. This means that classes of graphs exist where all known polynomial-time algorithms fail, in which case stochastic algorithms may possibly be preferable.

We show how GBS can be used to enhance stochastic algorithms. These algorithms combine exploration of the problem space with exploitation of local structure. Exploration can be achieved by randomly searching through the space, while exploitation involves tweaking candidate solutions and checking for an improvement. For graph problems, tweaking can be an operation where a candidate subgraph is modified by replacing a random subset of its vertices with other randomly chosen vertices. Classical algorithms employ uniform randomness for exploration and exploitation. However, following Ref.~\cite{arrazola2018quantum_published}, we can use biased randomness from GBS to enhance stochastic algorithms for the D$k$S problem. Crucially, this improvement is not algorithm-specific and works for any method using exploration and exploitation, regardless of inner details of the routine.

To enhance exploration, one simply samples from the GBS distribution of Eq. \eqref{Eq: GBS Graphs}, as formalised by the routine \textsf{GBS-Explore} in Ref.~\cite{arrazola2018quantum_published}. For exploitation, we can improve the tweak stage by using GBS to randomly select which vertices of candidate subgraphs to remove and also which ones to replace them with. More precisely, for a subgraph of even $k$ vertices with adjacency matrix $A_S$, we perform the following routine \textsf{GBS-Tweak} for a fixed even $l<k$ denoting the minimum number of vertices to be left untweaked:
\begin{enumerate}
\item Generate $R$ as an $l$ vertex subgraph of $S$ with adjacency matrix $A_{S,R}$ according to the GBS distribution $P_{l\mbox{cf}} \sim |\Haf (A_{S,R})|^{2}$. Extend $R$ by picking a uniform random number $m \in \{0,1,\ldots, k - l - 1\}$ of the vertices remaining from $S$, along with the corresponding edges. This is the subgraph $A_{S,keep}$ that specifies the $l+m$ vertices to be kept.
\item Generate $T$ as a $k-l$ vertex subgraph of $A$ with adjacency matrix $A_{T}$ according to the GBS distribution $P_{(k-l)\mbox{cf}} \sim |\Haf (A_{T})|^{2}$. Reduce $T$ by randomly rejecting $m$ of its vertices and corresponding edges. This is the subgraph $A_{T,replace}$ that specifies the $k-l-m$ vertices that will be added to $A_{S,keep}$. If $A_{S,keep}$ and $A_{T,replace}$ share any vertices, repeat this step.
\item Output the $k$ vertex subgraph $A_{S,keep} \bigcup A_{T,replace}$.
\end{enumerate}
GBS allows tweaking itself to become exploitative, with a two-fold improvement: since $R$ and $T$ are likely to be dense subgraphs, their composition should also be dense. We introduce the random parameter $m$ to vary the number of tweaked vertices.

\begin{figure*}[t!]
\begin{center}
\includegraphics[width= 0.93\columnwidth]{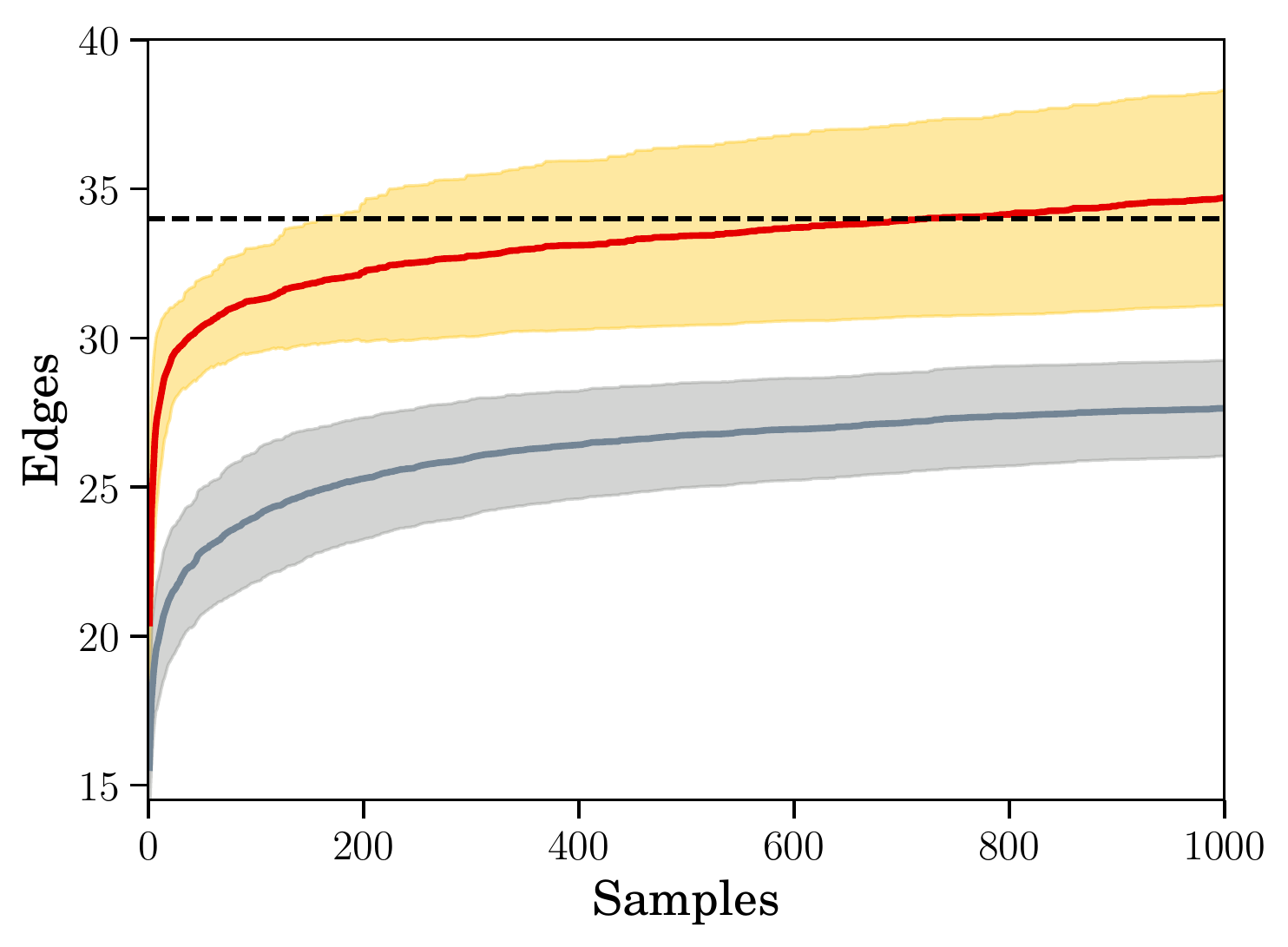}
\includegraphics[width= 0.93\columnwidth]{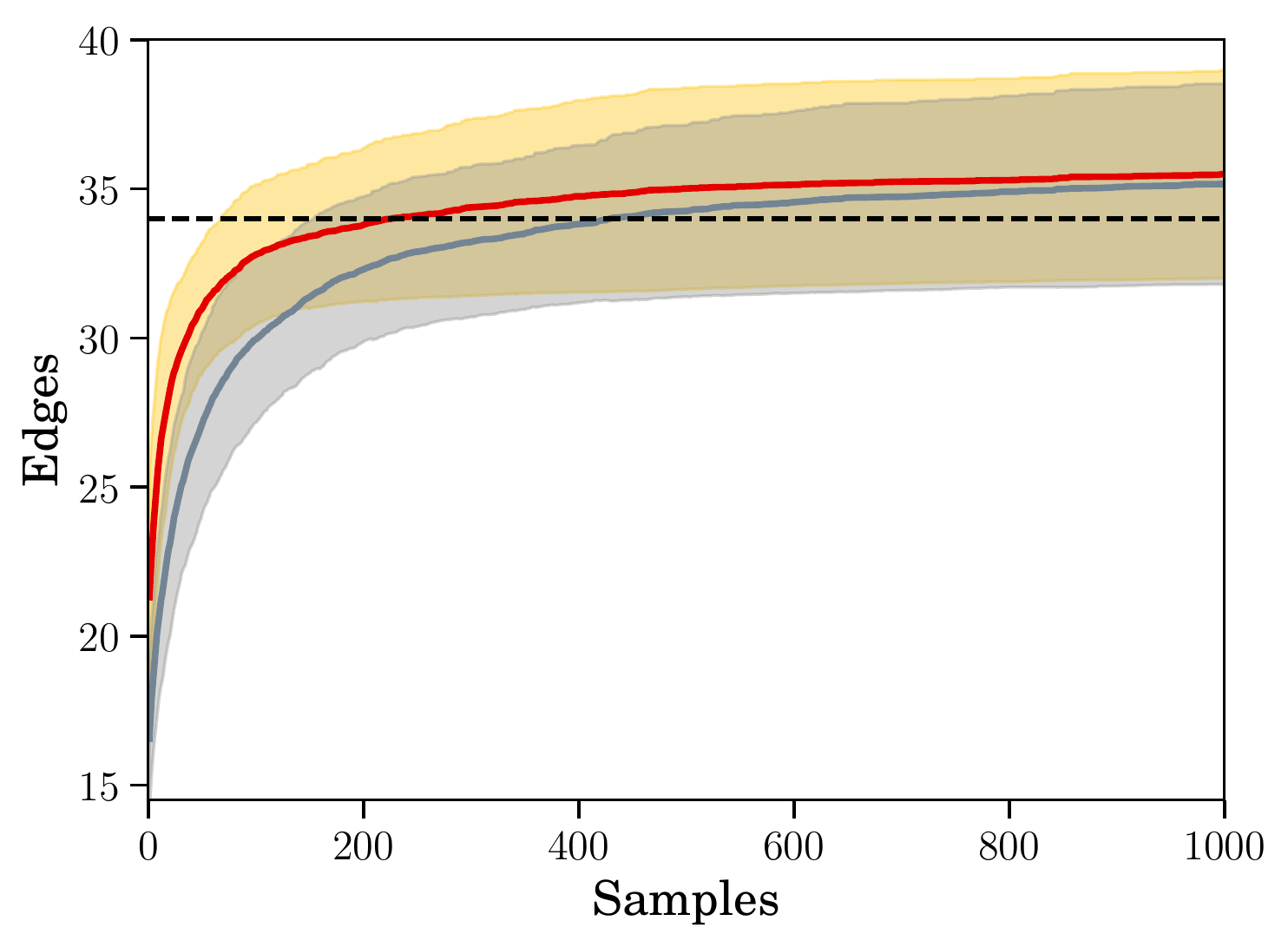}
\caption{Performance of random search (left) and simulated annealing (right) algorithms for finding the densest subgraph of 10 vertices in the graph of Fig. \ref{Fig: Graph}. The top red curve corresponds to using using GBS exploration and (for simulated annealing) GBS tweaking, while the bottom grey curve corresponds to the uniform random counterparts. The solid curves are the averages over $400$ repetitions and the error bars represent one standard deviation. The straight horizontal line shows the number of edges, $34$, in the dense subgraph found by the algorithm of Ref. \cite{charikar2000greedy}. The densest subgraph has 42 edges.}\label{Fig:Numerics}
\end{center}
\end{figure*}

\begin{center}
\begin{figure}[t!]
\includegraphics[width=0.65 \columnwidth]{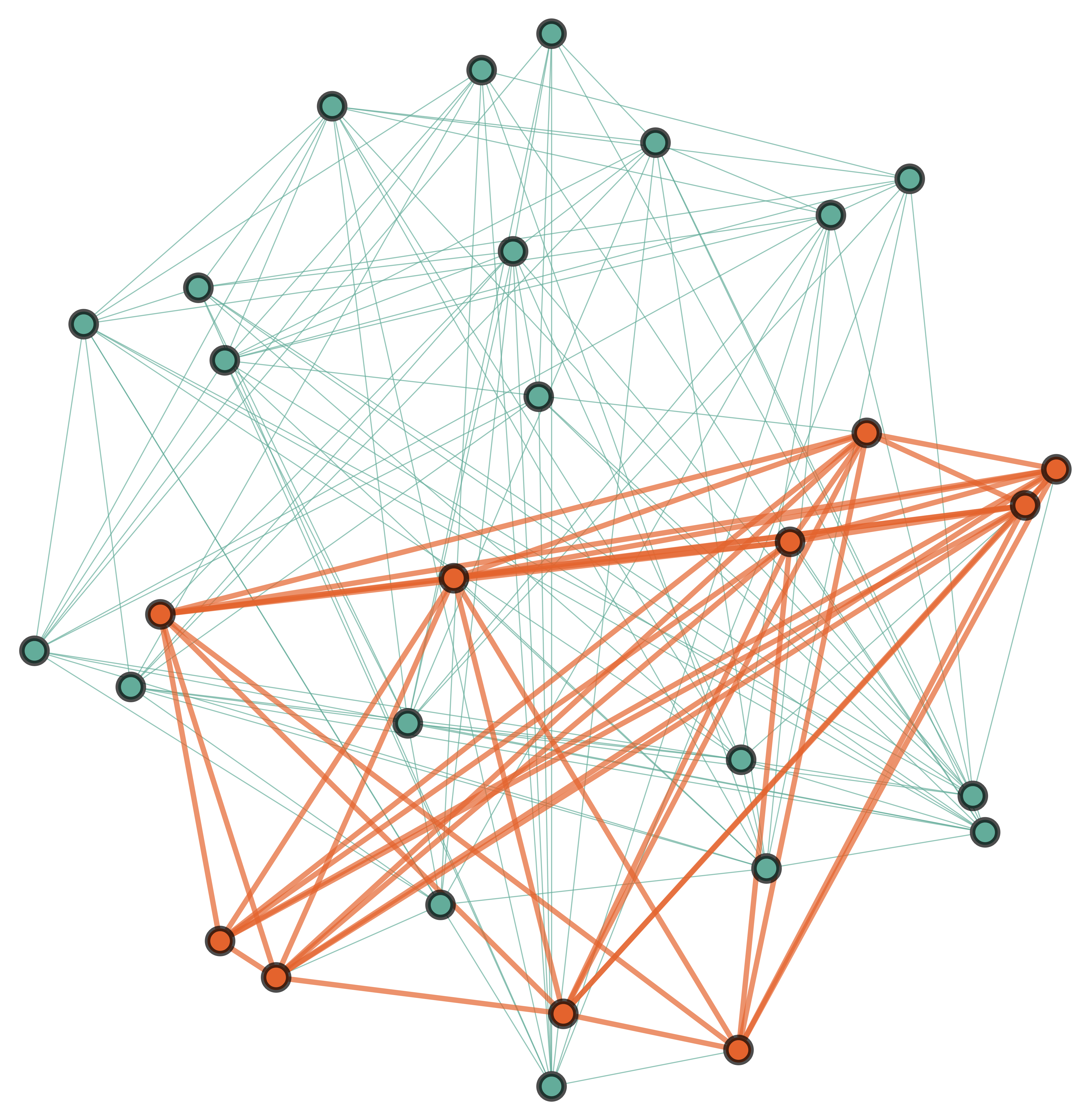}
\caption{A random graph of 30 vertices with a planted dense subgraph of 10 vertices (highlighted in red with thick edges). Vertices in the planted subgraph have lower degree than most other vertices in the graph, yet the density of the planted subgraph is the highest. This property prevents degree-based deterministic algorithms from finding the planted subgraph. } \label{Fig: Graph}
\end{figure}
\end{center}

GBS enhanced exploration and exploitation can be used within stochastic algorithms. Since random search only uses exploration, we discuss another example here. Simulated annealing is a heuristic optimization algorithm that combines elements of random search and hill climbing~\cite{vanLaarhoven1987simulated}. Whenever a new subgraph is generated, if its density is larger than the current one, it is retained. If its density is smaller, the new submatrix can still be retained with a probability that depends on the difference between the densities and a temperature parameter. The temperature is initially high and new subgraphs are often accepted, even if they have lower density. This is a feature that can prevent the search from becoming stuck in local minima. As the algorithm progresses, the temperature is lowered and only denser submatrices are kept, leading to an effective hill-climbing behavior. This algorithm is detailed in pseudocode in the Appendix.

\textit{Example D$k$S problem.---} To illustrate the enhancement to stochastic algorithms provided by GBS, we apply GBS enhanced random search and simulated annealing to the problem of locating a planted subgraph with large density, but whose vertices have low degree compared to the rest of the graph, see Fig.~\ref{Fig: Graph}. Such low-degree planted graphs are, by construction, hard to find for deterministic algorithms based on vertex degree. These graphs can model the presence of tightly-knit but otherwise isolated communities in social networks: members of these communities are highly-connected to each other (large density) but have few connections in total compared to typical members of the broader network (low degree). Note that more advanced deterministic algorithms can be designed to solve the D$k$S problem for this family of graphs \cite{hazan2011hard}.

To access GBS samples, we use the Hafnian formula of Ref.~\cite{bjorklund2018faster} to perform a brute force simulation of the entire probability distribution, which limits the size of graphs that we can sample from. Our graph was fixed to $30$ vertices with a planted subgraph of $10$ vertices. The graph was constructed by (i) generating a random graph of 20 vertices with probability $p=0.5$ of adding an edge, (ii) creating a random subgraph of 10 vertices with probability $q=0.875$ of having an edge (iii) selecting 8 vertices at random in both graphs and adding an edge between them. The result is shown in Fig.~\ref{Fig: Graph}. Here, the planted vertices have a lower average degree than other vertices, leading to a planted graph that is invisible to algorithms based on vertex degree. 

Figure~\ref{Fig:Numerics} illustrates the performance of random search and simulated annealing.
The plots each show the results of using GBS and uniform sampling in explore and exploit stages. The results are averaged over $400$ repetitions to remove statistical fluctuations, with the standard deviation also included. The simulated annealing parameters are $T = 0.01$, with a linear cooling schedule, and $l=6$. Here it is relevant to compare both the performance of simulated annealing over random search and the performance of using GBS over uniform sampling. It is first clear to see that GBS provides an advantage in both cases, illustrating our general findings that GBS is an enhancement for stochastic optimization algorithms. Furthermore, we see that simulated annealing is typically superior to random sampling and extends earlier beyond the region accessible by the deterministic algorithm of Ref. \cite{charikar2000greedy} ($32$ edges). Note however that GBS random search is particularly successful in the low sample number regime, outperforming both uniform and GBS simulated annealing for less than $50$ samples. This is a remarkable observation given the simplicity of GBS random search.

\textit{Discussion.---} We have shown that Gaussian boson sampling (GBS) is a useful tool for finding dense subgraphs. This results from the capability of GBS to perform proportional sampling for the canonical problem known as Max-Haf, highlighted in Ref.~\cite{arrazola2018quantum_published}, as well as the link between the number of perfect matchings (given by the Hafnian) and the density of a graph. This allows for tailored stochastic algorithms to be constructed for finding approximate solutions to the densest $k$-subgraph (D$k$S) problem.

It is important to emphasize that in the context of optimization, GBS is best understood as a quantum enhancement of stochastic algorithms. Although accurate deterministic algorithms exist, they can always be fooled under certain circumstances. Indeed, the D$k$S problem is NP-Hard and there are difficult instances for which no polynomial-time approximation algorithms exist, assuming the exponential time hypothesis~\cite{manurangsi2017almost}. This highlights a situation where stochastic algorithms, and their enhancement through GBS, are expected to be useful. Note that well-performing deterministic algorithms may also be enhanced through GBS by designing randomized versions.

These findings move away from the traditional approach to constructing quantum algorithms of rigorously showing a speedup in comparison to the best known classical algorithms. The heuristics-based approach followed here can instead allow for quantum enhancements to be identified in near-term devices. Overall, further research is needed to fully understand the potential advantages of enhancing stochastic algorithms through GBS when compared to highly optimized classical deterministic algorithms for dense subgraph identification and approximate optimization in general.

\textit{Acknowledgements.---} The authors thank Alex Arkhipov, Kamil Br\'adler, Pierre-Luc Dallaire-Demers, Nathan Killoran, Seth Lloyd, Patrick Rebentrost, Christian Weedbrook, and an anonymous referee for valuable discussions.

\newpage

\appendix
\section*{Appendix}

\begin{algorithm}[H]
  \caption{GBS Simulated Annealing for the D$k$S problem}
\begin{algorithmic}
\STATE $A$ : Input $n\times n$ adjacency matrix
\STATE $k$ : Dimension of submatrix
\STATE $l$ : Minimum number of entries to change when tweaking
\STATE $t$ : Initial temperature
\STATE $a_{\text{max}}$ : Number of steps
\STATE $S =\textsf{GBS-Explore}(k,A)$\COMMENT{This generates a submatrix from the corresponding GBS distribution.}
\STATE Best = $S$
\STATE \textbf{for} $a$ from 1 to $a_{\text{max}}$:
\STATE $\hspace{5mm}$ $R = \textsf{GBS-Tweak}(S,l,A)$\COMMENT{This tweaks the submatrix by randomly replacing vertices using GBS.}
\STATE $\hspace{5mm}$ \textbf{if} Density($R$) $>$ Density($S$):
\STATE $\hspace{10mm}$ $S = R$
\STATE $\hspace{5mm}$ \textbf{else}:  
\STATE $\hspace{10mm}$ Set $S = R$ with probability $\exp[\frac{\text{Density}(R) - \text{Density}(S)}{t}]$
\STATE $\hspace{5mm}$ \textbf{if} Density($S$) $>$ Density(Best):
\STATE $\hspace{10mm}$ Best = $S$
\STATE $\hspace{5mm}$ Decrease $t$
\STATE \textbf{end for}
\STATE Output Best
\end{algorithmic}
\end{algorithm}


\begin{thebibliography}{48}
\expandafter\ifx\csname natexlab\endcsname\relax\def\natexlab#1{#1}\fi
\expandafter\ifx\csname bibnamefont\endcsname\relax
  \def\bibnamefont#1{#1}\fi
\expandafter\ifx\csname bibfnamefont\endcsname\relax
  \def\bibfnamefont#1{#1}\fi
\expandafter\ifx\csname citenamefont\endcsname\relax
  \def\citenamefont#1{#1}\fi
\expandafter\ifx\csname url\endcsname\relax
  \def\url#1{\texttt{#1}}\fi
\expandafter\ifx\csname urlprefix\endcsname\relax\def\urlprefix{URL }\fi
\providecommand{\bibinfo}[2]{#2}
\providecommand{\eprint}[2][]{\url{#2}}

\bibitem[{\citenamefont{Zhang et~al.}(2017)\citenamefont{Zhang, Pagano, Hess,
  Kyprianidis, Becker, Kaplan, Gorshkov, Gong, and
  Monroe}}]{zhang2017observation}
\bibinfo{author}{\bibfnamefont{J.}~\bibnamefont{Zhang}},
  \bibinfo{author}{\bibfnamefont{G.}~\bibnamefont{Pagano}},
  \bibinfo{author}{\bibfnamefont{P.~W.} \bibnamefont{Hess}},
  \bibinfo{author}{\bibfnamefont{A.}~\bibnamefont{Kyprianidis}},
  \bibinfo{author}{\bibfnamefont{P.}~\bibnamefont{Becker}},
  \bibinfo{author}{\bibfnamefont{H.}~\bibnamefont{Kaplan}},
  \bibinfo{author}{\bibfnamefont{A.~V.} \bibnamefont{Gorshkov}},
  \bibinfo{author}{\bibfnamefont{Z.-X.} \bibnamefont{Gong}}, \bibnamefont{and}
  \bibinfo{author}{\bibfnamefont{C.}~\bibnamefont{Monroe}},
  \bibinfo{journal}{Nature} \textbf{\bibinfo{volume}{551}},
  \bibinfo{pages}{601} (\bibinfo{year}{2017}).

\bibitem[{\citenamefont{Bernien et~al.}(2017)\citenamefont{Bernien, Schwartz,
  Keesling, Levine, Omran, Pichler, Choi, Zibrov, Endres, Greiner
  et~al.}}]{bernien2017probing}
\bibinfo{author}{\bibfnamefont{H.}~\bibnamefont{Bernien}},
  \bibinfo{author}{\bibfnamefont{S.}~\bibnamefont{Schwartz}},
  \bibinfo{author}{\bibfnamefont{A.}~\bibnamefont{Keesling}},
  \bibinfo{author}{\bibfnamefont{H.}~\bibnamefont{Levine}},
  \bibinfo{author}{\bibfnamefont{A.}~\bibnamefont{Omran}},
  \bibinfo{author}{\bibfnamefont{H.}~\bibnamefont{Pichler}},
  \bibinfo{author}{\bibfnamefont{S.}~\bibnamefont{Choi}},
  \bibinfo{author}{\bibfnamefont{A.~S.} \bibnamefont{Zibrov}},
  \bibinfo{author}{\bibfnamefont{M.}~\bibnamefont{Endres}},
  \bibinfo{author}{\bibfnamefont{M.}~\bibnamefont{Greiner}},
  \bibnamefont{et~al.}, \bibinfo{journal}{Nature}
  \textbf{\bibinfo{volume}{551}}, \bibinfo{pages}{579} (\bibinfo{year}{2017}).

\bibitem[{\citenamefont{Peruzzo et~al.}(2014)\citenamefont{Peruzzo, McClean,
  Shadbolt, Yung, Zhou, Love, Aspuru-Guzik, and
  O'Brien}}]{peruzzo2014variational}
\bibinfo{author}{\bibfnamefont{A.}~\bibnamefont{Peruzzo}},
  \bibinfo{author}{\bibfnamefont{J.}~\bibnamefont{McClean}},
  \bibinfo{author}{\bibfnamefont{P.}~\bibnamefont{Shadbolt}},
  \bibinfo{author}{\bibfnamefont{M.-H.} \bibnamefont{Yung}},
  \bibinfo{author}{\bibfnamefont{X.-Q.} \bibnamefont{Zhou}},
  \bibinfo{author}{\bibfnamefont{P.~J.} \bibnamefont{Love}},
  \bibinfo{author}{\bibfnamefont{A.}~\bibnamefont{Aspuru-Guzik}},
  \bibnamefont{and} \bibinfo{author}{\bibfnamefont{J.~L.}
  \bibnamefont{O'Brien}}, \bibinfo{journal}{Nature Communications}
  \textbf{\bibinfo{volume}{5}}, \bibinfo{pages}{4213} (\bibinfo{year}{2014}).

\bibitem[{\citenamefont{McClean et~al.}(2016)\citenamefont{McClean, Romero,
  Babbush, and Aspuru-Guzik}}]{mcclean2016theory}
\bibinfo{author}{\bibfnamefont{J.~R.} \bibnamefont{McClean}},
  \bibinfo{author}{\bibfnamefont{J.}~\bibnamefont{Romero}},
  \bibinfo{author}{\bibfnamefont{R.}~\bibnamefont{Babbush}}, \bibnamefont{and}
  \bibinfo{author}{\bibfnamefont{A.}~\bibnamefont{Aspuru-Guzik}},
  \bibinfo{journal}{New Journal of Physics} \textbf{\bibinfo{volume}{18}},
  \bibinfo{pages}{023023} (\bibinfo{year}{2016}).

\bibitem[{\citenamefont{Moll et~al.}(2017)\citenamefont{Moll, Barkoutsos,
  Bishop, Chow, Cross, Egger, Filipp, Fuhrer, Gambetta, Ganzhorn
  et~al.}}]{moll2017quantum}
\bibinfo{author}{\bibfnamefont{N.}~\bibnamefont{Moll}},
  \bibinfo{author}{\bibfnamefont{P.}~\bibnamefont{Barkoutsos}},
  \bibinfo{author}{\bibfnamefont{L.~S.} \bibnamefont{Bishop}},
  \bibinfo{author}{\bibfnamefont{J.~M.} \bibnamefont{Chow}},
  \bibinfo{author}{\bibfnamefont{A.}~\bibnamefont{Cross}},
  \bibinfo{author}{\bibfnamefont{D.~J.} \bibnamefont{Egger}},
  \bibinfo{author}{\bibfnamefont{S.}~\bibnamefont{Filipp}},
  \bibinfo{author}{\bibfnamefont{A.}~\bibnamefont{Fuhrer}},
  \bibinfo{author}{\bibfnamefont{J.~M.} \bibnamefont{Gambetta}},
  \bibinfo{author}{\bibfnamefont{M.}~\bibnamefont{Ganzhorn}},
  \bibnamefont{et~al.}, \bibinfo{journal}{arXiv:1710.01022}
  (\bibinfo{year}{2017}).

\bibitem[{\citenamefont{Kandala et~al.}(2017)\citenamefont{Kandala, Mezzacapo,
  Temme, Takita, Brink, Chow, and Gambetta}}]{kandala2017hardware}
\bibinfo{author}{\bibfnamefont{A.}~\bibnamefont{Kandala}},
  \bibinfo{author}{\bibfnamefont{A.}~\bibnamefont{Mezzacapo}},
  \bibinfo{author}{\bibfnamefont{K.}~\bibnamefont{Temme}},
  \bibinfo{author}{\bibfnamefont{M.}~\bibnamefont{Takita}},
  \bibinfo{author}{\bibfnamefont{M.}~\bibnamefont{Brink}},
  \bibinfo{author}{\bibfnamefont{J.~M.} \bibnamefont{Chow}}, \bibnamefont{and}
  \bibinfo{author}{\bibfnamefont{J.~M.} \bibnamefont{Gambetta}},
  \bibinfo{journal}{Nature} \textbf{\bibinfo{volume}{549}},
  \bibinfo{pages}{242} (\bibinfo{year}{2017}).

\bibitem[{\citenamefont{Farhi et~al.}(2014)\citenamefont{Farhi, Goldstone, and
  Gutmann}}]{farhi2014quantum}
\bibinfo{author}{\bibfnamefont{E.}~\bibnamefont{Farhi}},
  \bibinfo{author}{\bibfnamefont{J.}~\bibnamefont{Goldstone}},
  \bibnamefont{and} \bibinfo{author}{\bibfnamefont{S.}~\bibnamefont{Gutmann}},
  \bibinfo{journal}{arXiv:1411.4028}  (\bibinfo{year}{2014}).

\bibitem[{\citenamefont{Farhi and Harrow}(2016)}]{farhi2016quantum}
\bibinfo{author}{\bibfnamefont{E.}~\bibnamefont{Farhi}} \bibnamefont{and}
  \bibinfo{author}{\bibfnamefont{A.~W.} \bibnamefont{Harrow}},
  \bibinfo{journal}{arXiv:1602.07674}  (\bibinfo{year}{2016}).

\bibitem[{\citenamefont{Li et~al.}(2015)\citenamefont{Li, Liu, Xu, and
  Du}}]{li2015experimental}
\bibinfo{author}{\bibfnamefont{Z.}~\bibnamefont{Li}},
  \bibinfo{author}{\bibfnamefont{X.}~\bibnamefont{Liu}},
  \bibinfo{author}{\bibfnamefont{N.}~\bibnamefont{Xu}}, \bibnamefont{and}
  \bibinfo{author}{\bibfnamefont{J.}~\bibnamefont{Du}},
  \bibinfo{journal}{Physical Review Letters} \textbf{\bibinfo{volume}{114}},
  \bibinfo{pages}{140504} (\bibinfo{year}{2015}).

\bibitem[{\citenamefont{Rist{\`e} et~al.}(2017)\citenamefont{Rist{\`e},
  Da~Silva, Ryan, Cross, C{\'o}rcoles, Smolin, Gambetta, Chow, and
  Johnson}}]{riste2017demonstration}
\bibinfo{author}{\bibfnamefont{D.}~\bibnamefont{Rist{\`e}}},
  \bibinfo{author}{\bibfnamefont{M.~P.} \bibnamefont{Da~Silva}},
  \bibinfo{author}{\bibfnamefont{C.~A.} \bibnamefont{Ryan}},
  \bibinfo{author}{\bibfnamefont{A.~W.} \bibnamefont{Cross}},
  \bibinfo{author}{\bibfnamefont{A.~D.} \bibnamefont{C{\'o}rcoles}},
  \bibinfo{author}{\bibfnamefont{J.~A.} \bibnamefont{Smolin}},
  \bibinfo{author}{\bibfnamefont{J.~M.} \bibnamefont{Gambetta}},
  \bibinfo{author}{\bibfnamefont{J.~M.} \bibnamefont{Chow}}, \bibnamefont{and}
  \bibinfo{author}{\bibfnamefont{B.~R.} \bibnamefont{Johnson}},
  \bibinfo{journal}{npj Quantum Information} \textbf{\bibinfo{volume}{3}},
  \bibinfo{pages}{16} (\bibinfo{year}{2017}).

\bibitem[{\citenamefont{Benedetti et~al.}(2017)\citenamefont{Benedetti,
  Realpe-G{\'o}mez, Biswas, and Perdomo-Ortiz}}]{benedetti2017quantum}
\bibinfo{author}{\bibfnamefont{M.}~\bibnamefont{Benedetti}},
  \bibinfo{author}{\bibfnamefont{J.}~\bibnamefont{Realpe-G{\'o}mez}},
  \bibinfo{author}{\bibfnamefont{R.}~\bibnamefont{Biswas}}, \bibnamefont{and}
  \bibinfo{author}{\bibfnamefont{A.}~\bibnamefont{Perdomo-Ortiz}},
  \bibinfo{journal}{Physical Review X} \textbf{\bibinfo{volume}{7}},
  \bibinfo{pages}{041052} (\bibinfo{year}{2017}).

\bibitem[{\citenamefont{Otterbach et~al.}(2017)\citenamefont{Otterbach,
  Manenti, Alidoust, Bestwick, Block, Bloom, Caldwell, Didier, Fried, Hong
  et~al.}}]{otterbach2017unsupervised}
\bibinfo{author}{\bibfnamefont{J.}~\bibnamefont{Otterbach}},
  \bibinfo{author}{\bibfnamefont{R.}~\bibnamefont{Manenti}},
  \bibinfo{author}{\bibfnamefont{N.}~\bibnamefont{Alidoust}},
  \bibinfo{author}{\bibfnamefont{A.}~\bibnamefont{Bestwick}},
  \bibinfo{author}{\bibfnamefont{M.}~\bibnamefont{Block}},
  \bibinfo{author}{\bibfnamefont{B.}~\bibnamefont{Bloom}},
  \bibinfo{author}{\bibfnamefont{S.}~\bibnamefont{Caldwell}},
  \bibinfo{author}{\bibfnamefont{N.}~\bibnamefont{Didier}},
  \bibinfo{author}{\bibfnamefont{E.~S.} \bibnamefont{Fried}},
  \bibinfo{author}{\bibfnamefont{S.}~\bibnamefont{Hong}}, \bibnamefont{et~al.},
  \bibinfo{journal}{arXiv:1712.05771}  (\bibinfo{year}{2017}).

\bibitem[{\citenamefont{Schuld and Killoran}(2018)}]{schuld2018quantum}
\bibinfo{author}{\bibfnamefont{M.}~\bibnamefont{Schuld}} \bibnamefont{and}
  \bibinfo{author}{\bibfnamefont{N.}~\bibnamefont{Killoran}},
  \bibinfo{journal}{arXiv:1803.07128}  (\bibinfo{year}{2018}).

\bibitem[{\citenamefont{Aaronson and
  Arkhipov}(2011)}]{aaronson2011computational}
\bibinfo{author}{\bibfnamefont{S.}~\bibnamefont{Aaronson}} \bibnamefont{and}
  \bibinfo{author}{\bibfnamefont{A.}~\bibnamefont{Arkhipov}}, in
  \emph{\bibinfo{booktitle}{Proceedings of the forty-third annual ACM symposium
  on Theory of computing}} (\bibinfo{organization}{ACM}, \bibinfo{address}{New
  York}, \bibinfo{year}{2011}), pp. \bibinfo{pages}{333--342}.

\bibitem[{\citenamefont{Spring et~al.}(2013)\citenamefont{Spring, Metcalf,
  Humphreys, Kolthammer, Jin, Barbieri, Datta, Thomas-Peter, Langford, Kundys
  et~al.}}]{spring1231692}
\bibinfo{author}{\bibfnamefont{J.~B.} \bibnamefont{Spring}},
  \bibinfo{author}{\bibfnamefont{B.~J.} \bibnamefont{Metcalf}},
  \bibinfo{author}{\bibfnamefont{P.~C.} \bibnamefont{Humphreys}},
  \bibinfo{author}{\bibfnamefont{W.~S.} \bibnamefont{Kolthammer}},
  \bibinfo{author}{\bibfnamefont{X.-M.} \bibnamefont{Jin}},
  \bibinfo{author}{\bibfnamefont{M.}~\bibnamefont{Barbieri}},
  \bibinfo{author}{\bibfnamefont{A.}~\bibnamefont{Datta}},
  \bibinfo{author}{\bibfnamefont{N.}~\bibnamefont{Thomas-Peter}},
  \bibinfo{author}{\bibfnamefont{N.~K.} \bibnamefont{Langford}},
  \bibinfo{author}{\bibfnamefont{D.}~\bibnamefont{Kundys}},
  \bibnamefont{et~al.}, \bibinfo{journal}{Science}
  \textbf{\bibinfo{volume}{339}}, \bibinfo{pages}{798} (\bibinfo{year}{2013}).

\bibitem[{\citenamefont{Broome et~al.}(2013)\citenamefont{Broome, Fedrizzi,
  Rahimi-Keshari, Dove, Aaronson, Ralph, and White}}]{broome794}
\bibinfo{author}{\bibfnamefont{M.~A.} \bibnamefont{Broome}},
  \bibinfo{author}{\bibfnamefont{A.}~\bibnamefont{Fedrizzi}},
  \bibinfo{author}{\bibfnamefont{S.}~\bibnamefont{Rahimi-Keshari}},
  \bibinfo{author}{\bibfnamefont{J.}~\bibnamefont{Dove}},
  \bibinfo{author}{\bibfnamefont{S.}~\bibnamefont{Aaronson}},
  \bibinfo{author}{\bibfnamefont{T.~C.} \bibnamefont{Ralph}}, \bibnamefont{and}
  \bibinfo{author}{\bibfnamefont{A.~G.} \bibnamefont{White}},
  \bibinfo{journal}{Science} \textbf{\bibinfo{volume}{339}},
  \bibinfo{pages}{794} (\bibinfo{year}{2013}).

\bibitem[{\citenamefont{Tillmann et~al.}(2013)\citenamefont{Tillmann,
  Daki{\'c}, Heilmann, Nolte, Szameit, and Walther}}]{tillmann2013experimental}
\bibinfo{author}{\bibfnamefont{M.}~\bibnamefont{Tillmann}},
  \bibinfo{author}{\bibfnamefont{B.}~\bibnamefont{Daki{\'c}}},
  \bibinfo{author}{\bibfnamefont{R.}~\bibnamefont{Heilmann}},
  \bibinfo{author}{\bibfnamefont{S.}~\bibnamefont{Nolte}},
  \bibinfo{author}{\bibfnamefont{A.}~\bibnamefont{Szameit}}, \bibnamefont{and}
  \bibinfo{author}{\bibfnamefont{P.}~\bibnamefont{Walther}},
  \bibinfo{journal}{Nature Photonics} \textbf{\bibinfo{volume}{7}},
  \bibinfo{pages}{540} (\bibinfo{year}{2013}).

\bibitem[{\citenamefont{Crespi et~al.}(2013)\citenamefont{Crespi, Osellame,
  Ramponi, Brod, Galvao, Spagnolo, Vitelli, Maiorino, Mataloni, and
  Sciarrino}}]{crespi2013integrated}
\bibinfo{author}{\bibfnamefont{A.}~\bibnamefont{Crespi}},
  \bibinfo{author}{\bibfnamefont{R.}~\bibnamefont{Osellame}},
  \bibinfo{author}{\bibfnamefont{R.}~\bibnamefont{Ramponi}},
  \bibinfo{author}{\bibfnamefont{D.~J.} \bibnamefont{Brod}},
  \bibinfo{author}{\bibfnamefont{E.~F.} \bibnamefont{Galvao}},
  \bibinfo{author}{\bibfnamefont{N.}~\bibnamefont{Spagnolo}},
  \bibinfo{author}{\bibfnamefont{C.}~\bibnamefont{Vitelli}},
  \bibinfo{author}{\bibfnamefont{E.}~\bibnamefont{Maiorino}},
  \bibinfo{author}{\bibfnamefont{P.}~\bibnamefont{Mataloni}}, \bibnamefont{and}
  \bibinfo{author}{\bibfnamefont{F.}~\bibnamefont{Sciarrino}},
  \bibinfo{journal}{Nature Photonics} \textbf{\bibinfo{volume}{7}},
  \bibinfo{pages}{545} (\bibinfo{year}{2013}).

\bibitem[{\citenamefont{Lund et~al.}(2014)\citenamefont{Lund, Laing,
  Rahimi-Keshari, Rudolph, O'Brien, and Ralph}}]{lund2014boson}
\bibinfo{author}{\bibfnamefont{A.}~\bibnamefont{Lund}},
  \bibinfo{author}{\bibfnamefont{A.}~\bibnamefont{Laing}},
  \bibinfo{author}{\bibfnamefont{S.}~\bibnamefont{Rahimi-Keshari}},
  \bibinfo{author}{\bibfnamefont{T.}~\bibnamefont{Rudolph}},
  \bibinfo{author}{\bibfnamefont{J.~L.} \bibnamefont{O'Brien}},
  \bibnamefont{and} \bibinfo{author}{\bibfnamefont{T.}~\bibnamefont{Ralph}},
  \bibinfo{journal}{Physical Review Letters} \textbf{\bibinfo{volume}{113}},
  \bibinfo{pages}{100502} (\bibinfo{year}{2014}).

\bibitem[{\citenamefont{Bentivegna et~al.}(2015)\citenamefont{Bentivegna,
  Spagnolo, Vitelli, Flamini, Viggianiello, Latmiral, Mataloni, Brod,
  Galv{\~a}o, Crespi et~al.}}]{bentivegna2015experimental}
\bibinfo{author}{\bibfnamefont{M.}~\bibnamefont{Bentivegna}},
  \bibinfo{author}{\bibfnamefont{N.}~\bibnamefont{Spagnolo}},
  \bibinfo{author}{\bibfnamefont{C.}~\bibnamefont{Vitelli}},
  \bibinfo{author}{\bibfnamefont{F.}~\bibnamefont{Flamini}},
  \bibinfo{author}{\bibfnamefont{N.}~\bibnamefont{Viggianiello}},
  \bibinfo{author}{\bibfnamefont{L.}~\bibnamefont{Latmiral}},
  \bibinfo{author}{\bibfnamefont{P.}~\bibnamefont{Mataloni}},
  \bibinfo{author}{\bibfnamefont{D.~J.} \bibnamefont{Brod}},
  \bibinfo{author}{\bibfnamefont{E.~F.} \bibnamefont{Galv{\~a}o}},
  \bibinfo{author}{\bibfnamefont{A.}~\bibnamefont{Crespi}},
  \bibnamefont{et~al.}, \bibinfo{journal}{Science Advances}
  \textbf{\bibinfo{volume}{1}}, \bibinfo{pages}{e1400255}
  (\bibinfo{year}{2015}).

\bibitem[{\citenamefont{Latmiral et~al.}(2016)\citenamefont{Latmiral, Spagnolo,
  and Sciarrino}}]{latmiral2016towards}
\bibinfo{author}{\bibfnamefont{L.}~\bibnamefont{Latmiral}},
  \bibinfo{author}{\bibfnamefont{N.}~\bibnamefont{Spagnolo}}, \bibnamefont{and}
  \bibinfo{author}{\bibfnamefont{F.}~\bibnamefont{Sciarrino}},
  \bibinfo{journal}{New Journal of Physics} \textbf{\bibinfo{volume}{18}},
  \bibinfo{pages}{113008} (\bibinfo{year}{2016}).

\bibitem[{\citenamefont{Hamilton et~al.}(2017)\citenamefont{Hamilton, Kruse,
  Sansoni, Barkhofen, Silberhorn, and Jex}}]{hamilton2017gaussian}
\bibinfo{author}{\bibfnamefont{C.~S.} \bibnamefont{Hamilton}},
  \bibinfo{author}{\bibfnamefont{R.}~\bibnamefont{Kruse}},
  \bibinfo{author}{\bibfnamefont{L.}~\bibnamefont{Sansoni}},
  \bibinfo{author}{\bibfnamefont{S.}~\bibnamefont{Barkhofen}},
  \bibinfo{author}{\bibfnamefont{C.}~\bibnamefont{Silberhorn}},
  \bibnamefont{and} \bibinfo{author}{\bibfnamefont{I.}~\bibnamefont{Jex}},
  \bibinfo{journal}{Physical Review Letters} \textbf{\bibinfo{volume}{119}},
  \bibinfo{pages}{170501} (\bibinfo{year}{2017}).

\bibitem[{\citenamefont{Kruse et~al.}(2018)\citenamefont{Kruse, Hamilton,
  Sansoni, Barkhofen, Silberhorn, and Jex}}]{kruse2018detailed}
\bibinfo{author}{\bibfnamefont{R.}~\bibnamefont{Kruse}},
  \bibinfo{author}{\bibfnamefont{C.~S.} \bibnamefont{Hamilton}},
  \bibinfo{author}{\bibfnamefont{L.}~\bibnamefont{Sansoni}},
  \bibinfo{author}{\bibfnamefont{S.}~\bibnamefont{Barkhofen}},
  \bibinfo{author}{\bibfnamefont{C.}~\bibnamefont{Silberhorn}},
  \bibnamefont{and} \bibinfo{author}{\bibfnamefont{I.}~\bibnamefont{Jex}},
  \bibinfo{journal}{arXiv:1801.07488}  (\bibinfo{year}{2018}).

\bibitem[{\citenamefont{Huh et~al.}(2015)\citenamefont{Huh, Guerreschi,
  Peropadre, McClean, and Aspuru-Guzik}}]{huh2015boson}
\bibinfo{author}{\bibfnamefont{J.}~\bibnamefont{Huh}},
  \bibinfo{author}{\bibfnamefont{G.~G.} \bibnamefont{Guerreschi}},
  \bibinfo{author}{\bibfnamefont{B.}~\bibnamefont{Peropadre}},
  \bibinfo{author}{\bibfnamefont{J.~R.} \bibnamefont{McClean}},
  \bibnamefont{and}
  \bibinfo{author}{\bibfnamefont{A.}~\bibnamefont{Aspuru-Guzik}},
  \bibinfo{journal}{Nature Photonics} \textbf{\bibinfo{volume}{9}},
  \bibinfo{pages}{615} (\bibinfo{year}{2015}).

\bibitem[{\citenamefont{Clements et~al.}(2017)\citenamefont{Clements, Renema,
  Eckstein, Valido, Lita, Gerrits, Nam, Kolthammer, Huh, and
  Walmsley}}]{clements2017experimental}
\bibinfo{author}{\bibfnamefont{W.~R.} \bibnamefont{Clements}},
  \bibinfo{author}{\bibfnamefont{J.~J.} \bibnamefont{Renema}},
  \bibinfo{author}{\bibfnamefont{A.}~\bibnamefont{Eckstein}},
  \bibinfo{author}{\bibfnamefont{A.~A.} \bibnamefont{Valido}},
  \bibinfo{author}{\bibfnamefont{A.}~\bibnamefont{Lita}},
  \bibinfo{author}{\bibfnamefont{T.}~\bibnamefont{Gerrits}},
  \bibinfo{author}{\bibfnamefont{S.~W.} \bibnamefont{Nam}},
  \bibinfo{author}{\bibfnamefont{W.~S.} \bibnamefont{Kolthammer}},
  \bibinfo{author}{\bibfnamefont{J.}~\bibnamefont{Huh}}, \bibnamefont{and}
  \bibinfo{author}{\bibfnamefont{I.~A.} \bibnamefont{Walmsley}},
  \bibinfo{journal}{arXiv:1710.08655}  (\bibinfo{year}{2017}).

\bibitem[{\citenamefont{Sparrow et~al.}(2018)\citenamefont{Sparrow,
  Mart{\'\i}n-L{\'o}pez, Maraviglia, Neville, Harrold, Carolan, Joglekar,
  Hashimoto, Matsuda, O’Brien et~al.}}]{sparrow2018simulating}
\bibinfo{author}{\bibfnamefont{C.}~\bibnamefont{Sparrow}},
  \bibinfo{author}{\bibfnamefont{E.}~\bibnamefont{Mart{\'\i}n-L{\'o}pez}},
  \bibinfo{author}{\bibfnamefont{N.}~\bibnamefont{Maraviglia}},
  \bibinfo{author}{\bibfnamefont{A.}~\bibnamefont{Neville}},
  \bibinfo{author}{\bibfnamefont{C.}~\bibnamefont{Harrold}},
  \bibinfo{author}{\bibfnamefont{J.}~\bibnamefont{Carolan}},
  \bibinfo{author}{\bibfnamefont{Y.~N.} \bibnamefont{Joglekar}},
  \bibinfo{author}{\bibfnamefont{T.}~\bibnamefont{Hashimoto}},
  \bibinfo{author}{\bibfnamefont{N.}~\bibnamefont{Matsuda}},
  \bibinfo{author}{\bibfnamefont{J.~L.} \bibnamefont{O’Brien}},
  \bibnamefont{et~al.}, \bibinfo{journal}{Nature}
  \textbf{\bibinfo{volume}{557}}, \bibinfo{pages}{660} (\bibinfo{year}{2018}).

\bibitem[{\citenamefont{Feige et~al.}(2001)\citenamefont{Feige, Peleg, and
  Kortsarz}}]{feige2001dense}
\bibinfo{author}{\bibfnamefont{U.}~\bibnamefont{Feige}},
  \bibinfo{author}{\bibfnamefont{D.}~\bibnamefont{Peleg}}, \bibnamefont{and}
  \bibinfo{author}{\bibfnamefont{G.}~\bibnamefont{Kortsarz}},
  \bibinfo{journal}{Algorithmica} \textbf{\bibinfo{volume}{29}},
  \bibinfo{pages}{410} (\bibinfo{year}{2001}).

\bibitem[{\citenamefont{Kumar et~al.}(1999)\citenamefont{Kumar, Raghavan,
  Rajagopalan, and Tomkins}}]{kumar1999trawling}
\bibinfo{author}{\bibfnamefont{R.}~\bibnamefont{Kumar}},
  \bibinfo{author}{\bibfnamefont{P.}~\bibnamefont{Raghavan}},
  \bibinfo{author}{\bibfnamefont{S.}~\bibnamefont{Rajagopalan}},
  \bibnamefont{and} \bibinfo{author}{\bibfnamefont{A.}~\bibnamefont{Tomkins}},
  \bibinfo{journal}{Computer Networks} \textbf{\bibinfo{volume}{31}},
  \bibinfo{pages}{1481} (\bibinfo{year}{1999}).

\bibitem[{\citenamefont{Angel et~al.}(2012)\citenamefont{Angel, Sarkas, Koudas,
  and Srivastava}}]{angel2012dense}
\bibinfo{author}{\bibfnamefont{A.}~\bibnamefont{Angel}},
  \bibinfo{author}{\bibfnamefont{N.}~\bibnamefont{Sarkas}},
  \bibinfo{author}{\bibfnamefont{N.}~\bibnamefont{Koudas}}, \bibnamefont{and}
  \bibinfo{author}{\bibfnamefont{D.}~\bibnamefont{Srivastava}},
  \bibinfo{journal}{Proceedings of the VLDB Endowment}
  \textbf{\bibinfo{volume}{5}}, \bibinfo{pages}{574} (\bibinfo{year}{2012}).

\bibitem[{\citenamefont{Beutel et~al.}(2013)\citenamefont{Beutel, Xu,
  Guruswami, Palow, and Faloutsos}}]{beutel2013copycatch}
\bibinfo{author}{\bibfnamefont{A.}~\bibnamefont{Beutel}},
  \bibinfo{author}{\bibfnamefont{W.}~\bibnamefont{Xu}},
  \bibinfo{author}{\bibfnamefont{V.}~\bibnamefont{Guruswami}},
  \bibinfo{author}{\bibfnamefont{C.}~\bibnamefont{Palow}}, \bibnamefont{and}
  \bibinfo{author}{\bibfnamefont{C.}~\bibnamefont{Faloutsos}}, in
  \emph{\bibinfo{booktitle}{Proceedings of the 22nd international conference on
  World Wide Web}} (\bibinfo{organization}{ACM}, \bibinfo{address}{New York},
  \bibinfo{year}{2013}), pp. \bibinfo{pages}{119--130}.

\bibitem[{\citenamefont{Chen and Saad}(2012)}]{chen2012dense}
\bibinfo{author}{\bibfnamefont{J.}~\bibnamefont{Chen}} \bibnamefont{and}
  \bibinfo{author}{\bibfnamefont{Y.}~\bibnamefont{Saad}},
  \bibinfo{journal}{IEEE Transactions on Knowledge and Data Engineering}
  \textbf{\bibinfo{volume}{24}}, \bibinfo{pages}{1216} (\bibinfo{year}{2012}).

\bibitem[{\citenamefont{Fratkin et~al.}(2006)\citenamefont{Fratkin, Naughton,
  Brutlag, and Batzoglou}}]{fratkin2006motifcut}
\bibinfo{author}{\bibfnamefont{E.}~\bibnamefont{Fratkin}},
  \bibinfo{author}{\bibfnamefont{B.~T.} \bibnamefont{Naughton}},
  \bibinfo{author}{\bibfnamefont{D.~L.} \bibnamefont{Brutlag}},
  \bibnamefont{and}
  \bibinfo{author}{\bibfnamefont{S.}~\bibnamefont{Batzoglou}},
  \bibinfo{journal}{Bioinformatics} \textbf{\bibinfo{volume}{22}},
  \bibinfo{pages}{e150} (\bibinfo{year}{2006}).

\bibitem[{\citenamefont{Saha et~al.}(2010)\citenamefont{Saha, Hoch, Khuller,
  Raschid, and Zhang}}]{saha2010dense}
\bibinfo{author}{\bibfnamefont{B.}~\bibnamefont{Saha}},
  \bibinfo{author}{\bibfnamefont{A.}~\bibnamefont{Hoch}},
  \bibinfo{author}{\bibfnamefont{S.}~\bibnamefont{Khuller}},
  \bibinfo{author}{\bibfnamefont{L.}~\bibnamefont{Raschid}}, \bibnamefont{and}
  \bibinfo{author}{\bibfnamefont{X.-N.} \bibnamefont{Zhang}}, in
  \emph{\bibinfo{booktitle}{Annual International Conference on Research in
  Computational Molecular Biology}} (\bibinfo{organization}{Springer},
  \bibinfo{address}{Berlin}, \bibinfo{year}{2010}), pp.
  \bibinfo{pages}{456--472}.

\bibitem[{\citenamefont{Arora et~al.}(2011)\citenamefont{Arora, Barak,
  Brunnermeier, and Ge}}]{arora2011computational}
\bibinfo{author}{\bibfnamefont{S.}~\bibnamefont{Arora}},
  \bibinfo{author}{\bibfnamefont{B.}~\bibnamefont{Barak}},
  \bibinfo{author}{\bibfnamefont{M.}~\bibnamefont{Brunnermeier}},
  \bibnamefont{and} \bibinfo{author}{\bibfnamefont{R.}~\bibnamefont{Ge}},
  \bibinfo{journal}{Communications of the ACM} \textbf{\bibinfo{volume}{54}},
  \bibinfo{pages}{101} (\bibinfo{year}{2011}).

\bibitem[{\citenamefont{Br{\'a}dler et~al.}(2017)\citenamefont{Br{\'a}dler,
  Dallaire-Demers, Rebentrost, Su, and Weedbrook}}]{bradler2017gaussian}
\bibinfo{author}{\bibfnamefont{K.}~\bibnamefont{Br{\'a}dler}},
  \bibinfo{author}{\bibfnamefont{P.-L.} \bibnamefont{Dallaire-Demers}},
  \bibinfo{author}{\bibfnamefont{P.}~\bibnamefont{Rebentrost}},
  \bibinfo{author}{\bibfnamefont{D.}~\bibnamefont{Su}}, \bibnamefont{and}
  \bibinfo{author}{\bibfnamefont{C.}~\bibnamefont{Weedbrook}},
  \bibinfo{journal}{arXiv:1712.06729}  (\bibinfo{year}{2017}).

\bibitem[{\citenamefont{Arrazola et~al.}(2018)\citenamefont{Arrazola, Bromley,
  and Rebentrost}}]{arrazola2018quantum_published}
\bibinfo{author}{\bibfnamefont{J.~M.} \bibnamefont{Arrazola}},
  \bibinfo{author}{\bibfnamefont{T.~R.} \bibnamefont{Bromley}},
  \bibnamefont{and}
  \bibinfo{author}{\bibfnamefont{P.}~\bibnamefont{Rebentrost}},
  \bibinfo{journal}{Physical Review A} \textbf{\bibinfo{volume}{98}},
  \bibinfo{pages}{012322} (\bibinfo{year}{2018}).

\bibitem[{\citenamefont{Manurangsi}(2017)}]{manurangsi2017almost}
\bibinfo{author}{\bibfnamefont{P.}~\bibnamefont{Manurangsi}}, in
  \emph{\bibinfo{booktitle}{Proceedings of the 49th Annual ACM SIGACT Symposium
  on Theory of Computing}}, \bibinfo{organization}{ACM}
  (\bibinfo{publisher}{ACM}, \bibinfo{address}{Montreal},
  \bibinfo{year}{2017}), pp. \bibinfo{pages}{954--961}.

\bibitem[{\citenamefont{Weedbrook et~al.}(2012)\citenamefont{Weedbrook,
  Pirandola, Garc{\'\i}a-Patr{\'o}n, Cerf, Ralph, Shapiro, and
  Lloyd}}]{weedbrook2012gaussian}
\bibinfo{author}{\bibfnamefont{C.}~\bibnamefont{Weedbrook}},
  \bibinfo{author}{\bibfnamefont{S.}~\bibnamefont{Pirandola}},
  \bibinfo{author}{\bibfnamefont{R.}~\bibnamefont{Garc{\'\i}a-Patr{\'o}n}},
  \bibinfo{author}{\bibfnamefont{N.~J.} \bibnamefont{Cerf}},
  \bibinfo{author}{\bibfnamefont{T.~C.} \bibnamefont{Ralph}},
  \bibinfo{author}{\bibfnamefont{J.~H.} \bibnamefont{Shapiro}},
  \bibnamefont{and} \bibinfo{author}{\bibfnamefont{S.}~\bibnamefont{Lloyd}},
  \bibinfo{journal}{Reviews of Modern Physics} \textbf{\bibinfo{volume}{84}},
  \bibinfo{pages}{621} (\bibinfo{year}{2012}).

\bibitem[{\citenamefont{Barvinok}(2016)}]{barvinok2016combinatorics}
\bibinfo{author}{\bibfnamefont{A.}~\bibnamefont{Barvinok}},
  \emph{\bibinfo{title}{Combinatorics and complexity of partition functions}},
  vol. \bibinfo{volume}{274} (\bibinfo{publisher}{Springer},
  \bibinfo{address}{Berlin}, \bibinfo{year}{2016}).

\bibitem[{\citenamefont{Aaghabali et~al.}(2015)\citenamefont{Aaghabali, Akbari,
  Friedland, Markstr{\"o}m, and Tajfirouz}}]{aaghabali2015upper}
\bibinfo{author}{\bibfnamefont{M.}~\bibnamefont{Aaghabali}},
  \bibinfo{author}{\bibfnamefont{S.}~\bibnamefont{Akbari}},
  \bibinfo{author}{\bibfnamefont{S.}~\bibnamefont{Friedland}},
  \bibinfo{author}{\bibfnamefont{K.}~\bibnamefont{Markstr{\"o}m}},
  \bibnamefont{and}
  \bibinfo{author}{\bibfnamefont{Z.}~\bibnamefont{Tajfirouz}},
  \bibinfo{journal}{European Journal of Combinatorics}
  \textbf{\bibinfo{volume}{45}}, \bibinfo{pages}{132} (\bibinfo{year}{2015}).

\bibitem[{\citenamefont{Rudelson et~al.}(2016)\citenamefont{Rudelson,
  Samorodnitsky, and Zeitouni}}]{rudelson2016hafnians}
\bibinfo{author}{\bibfnamefont{M.}~\bibnamefont{Rudelson}},
  \bibinfo{author}{\bibfnamefont{A.}~\bibnamefont{Samorodnitsky}},
  \bibnamefont{and} \bibinfo{author}{\bibfnamefont{O.}~\bibnamefont{Zeitouni}},
  \bibinfo{journal}{The Annals of Probability} \textbf{\bibinfo{volume}{44}},
  \bibinfo{pages}{2858} (\bibinfo{year}{2016}).

\bibitem[{\citenamefont{Neville et~al.}(2017)\citenamefont{Neville, Sparrow,
  Clifford, Johnston, Birchall, Montanaro, and Laing}}]{neville2017classical}
\bibinfo{author}{\bibfnamefont{A.}~\bibnamefont{Neville}},
  \bibinfo{author}{\bibfnamefont{C.}~\bibnamefont{Sparrow}},
  \bibinfo{author}{\bibfnamefont{R.}~\bibnamefont{Clifford}},
  \bibinfo{author}{\bibfnamefont{E.}~\bibnamefont{Johnston}},
  \bibinfo{author}{\bibfnamefont{P.~M.} \bibnamefont{Birchall}},
  \bibinfo{author}{\bibfnamefont{A.}~\bibnamefont{Montanaro}},
  \bibnamefont{and} \bibinfo{author}{\bibfnamefont{A.}~\bibnamefont{Laing}},
  \bibinfo{journal}{Nature Physics} \textbf{\bibinfo{volume}{13}},
  \bibinfo{pages}{1153} (\bibinfo{year}{2017}).

\bibitem[{\citenamefont{Liu}(1996)}]{liu1996metropolized}
\bibinfo{author}{\bibfnamefont{J.~S.} \bibnamefont{Liu}},
  \bibinfo{journal}{Statistics and Computing} \textbf{\bibinfo{volume}{6}},
  \bibinfo{pages}{113} (\bibinfo{year}{1996}).

\bibitem[{\citenamefont{Lee et~al.}(2010)\citenamefont{Lee, Ruan, Jin, and
  Aggarwal}}]{lee2010survey}
\bibinfo{author}{\bibfnamefont{V.~E.} \bibnamefont{Lee}},
  \bibinfo{author}{\bibfnamefont{N.}~\bibnamefont{Ruan}},
  \bibinfo{author}{\bibfnamefont{R.}~\bibnamefont{Jin}}, \bibnamefont{and}
  \bibinfo{author}{\bibfnamefont{C.}~\bibnamefont{Aggarwal}}, in
  \emph{\bibinfo{booktitle}{Managing and Mining Graph Data}}
  (\bibinfo{publisher}{Springer}, \bibinfo{year}{2010}), pp.
  \bibinfo{pages}{303--336}.

\bibitem[{\citenamefont{Charikar}(2000)}]{charikar2000greedy}
\bibinfo{author}{\bibfnamefont{M.}~\bibnamefont{Charikar}}, in
  \emph{\bibinfo{booktitle}{International Workshop on Approximation Algorithms
  for Combinatorial Optimization}} (\bibinfo{organization}{Springer},
  \bibinfo{address}{Berlin}, \bibinfo{year}{2000}), pp.
  \bibinfo{pages}{84--95}.

\bibitem[{\citenamefont{van Laarhoven and
  Aarts}(1987)}]{vanLaarhoven1987simulated}
\bibinfo{author}{\bibfnamefont{P.}~\bibnamefont{van Laarhoven}}
  \bibnamefont{and} \bibinfo{author}{\bibfnamefont{E.}~\bibnamefont{Aarts}},
  \emph{\bibinfo{title}{Simulated Annealing: Theory and Applications}},
  vol.~\bibinfo{volume}{37} (\bibinfo{publisher}{Springer},
  \bibinfo{address}{Berlin}, \bibinfo{year}{1987}).

\bibitem[{\citenamefont{Hazan and Krauthgamer}(2011)}]{hazan2011hard}
\bibinfo{author}{\bibfnamefont{E.}~\bibnamefont{Hazan}} \bibnamefont{and}
  \bibinfo{author}{\bibfnamefont{R.}~\bibnamefont{Krauthgamer}},
  \bibinfo{journal}{SIAM Journal on Computing} \textbf{\bibinfo{volume}{40}},
  \bibinfo{pages}{79} (\bibinfo{year}{2011}).

\bibitem[{\citenamefont{Bj{\"o}rklund et~al.}(2018)\citenamefont{Bj{\"o}rklund,
  Gupt, and Quesada}}]{bjorklund2018faster}
\bibinfo{author}{\bibfnamefont{A.}~\bibnamefont{Bj{\"o}rklund}},
  \bibinfo{author}{\bibfnamefont{B.}~\bibnamefont{Gupt}}, \bibnamefont{and}
  \bibinfo{author}{\bibfnamefont{N.}~\bibnamefont{Quesada}},
  \bibinfo{journal}{arXiv preprint arXiv:1805.12498}  (\bibinfo{year}{2018}).

\end{thebibliography}
\end{document}